\documentclass[letterpaper,10pt]{article} 

\usepackage{opticameet3} 
\usepackage{caption}
\usepackage{changepage} 

\newcommand\authormark[1]{\textsuperscript{#1}}

\usepackage{amsmath,amssymb}
\usepackage[colorlinks=true,bookmarks=false,citecolor=blue,urlcolor=blue]{hyperref} 

\begin{document}

\begin{figure*}[b!]
    \centering
    \includegraphics[width=1\textwidth]{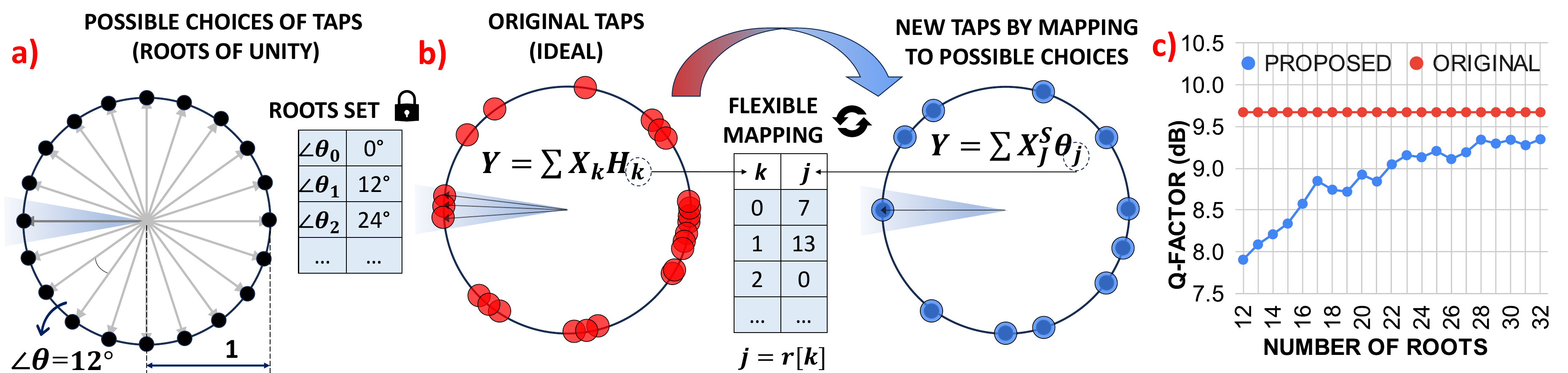}
    \captionsetup{font=footnotesize}
    \caption{a) The fixed set of possible choices for filter coefficients approximation (roots of unity). b)~Approximation and mapping scheme. c)~Performance for different numbers of roots at 16 spans (1280 km).}
    \label{root-analysis}
\end{figure*}

\title{

\begin{adjustwidth}{-1.5cm}{-1.5cm}  
\centering

FPGA Implementation of Low-Power Multiplierless \\ 
Pre-Processing Free Chromatic Dispersion Equalizer

\end{adjustwidth}
}





\vspace{-3mm}
\author{Geraldo Gomes,\authormark{1,*} Pedro Freire,\authormark{1,*}  Jaroslaw E. Prilepsky,\authormark{1} and Sergei K. Turitsyn\authormark{1}}

\address{\authormark{1} Aston University,  Aston Institute of Photonic Technologies, Birmingham, UK, B4 7ET\\
}

\email{\authormark{*}freiredp@aston.ac.uk} 

\vspace{-3mm}
\begin{abstract}
We present a novel time-domain chromatic dispersion equalizer, implemented on FPGA, eliminating pre-processing and multipliers, achieving up to 54.3\% energy savings over 80–1280 km with a simple, low-power design.
\end{abstract}

\section{Introduction}
Chromatic dispersion (CD) compensation continues to be one of the most power-hungry digital signal processing (DSP) blocks in modern coherent optical communication systems. The existing solutions often rely on FFT-based methods to handle CD, introducing significant complexity and power consumption. Here, we propose a novel, low-complexity, multiplierless, and pre-processing-free chromatic dispersion compensation (CDC) algorithm. 
Our approach, which we name the Roots of Unity Equalizer (RUE), approximates filter coefficients as equally spaced phase shifts with a constant amplitude of one (Fig. \ref{root-analysis}a), corresponding to the roots of unity ($z^n = 1$). Unlike FFT-based methods that involve computationally expensive multipliers and numerous operations for frequency-domain transformation, RUE operates entirely in the time domain using only bit shifts and adders, significantly reducing both complexity and power consumption. RUE also supports real-time dispersion estimation, eliminating the need for pre-processing, which is often a costly factor in real-world applications. Compared to FFT-based solutions, our FPGA implementation of RUE demonstrates up to 54.3\% energy savings, validated using the nJ/bit metric over fiber lengths ranging from 80 km to 1280 km.  Moreover, any scaling of the recovered constellation, due to the fixed amplitude, can be automatically corrected by the adaptive equalization already present in coherent transceivers \cite{savory2008digital}.

\section{Flexible and Multiplierless Implementation}

\subsection{Multiplierless Approach}

Using the geometric interpretation of CDC taps in the complex plane \cite{taylor2008compact,gomes2024fpgaimplementation}, we demonstrate a multiplierless, energy-efficient, and pre-processing-free implementation. Our method approximates the phase shifts of the  filter taps, $H_k$ (Fig.~\ref{root-analysis}b), by selecting from a fixed set of predefined roots (Fig.~\ref{root-analysis}a). This is achieved through a mapping strategy, as illustrated by the shaded areas, which groups several input samples to be multiplied by the same tap,~\( \theta_j \).

Applying the distributive property \cite{martins2016distributive}, the convolution can be rewritten as:
\(
Y = \sum X_j^S \cdot \theta_j,
\)
where $Y$ is the output sample, and \( X_j^S = \sum X_k \) represents the sum of input samples $X_k$ that are multiplied by the same complex filter tap \( \theta_j \), with the indices $j$ and $k$ determined by the mapping $j = r[k]$. Given that \( \theta_j = \theta_1^j \), this can be further simplified as: \(
Y = \sum X_j^S \cdot \theta_1^j,
\) which leads to a recursive formulation: \(
Y = X_0^S + \theta_1 (X_1^S + \theta_1 (X_2^S + \theta_1 (X_3^S + \dots)))
\) with all multiplications being performed using the same constant, \( \theta_1 \). The mapping $r[k]$ defines the number of shifts by the angle $\angle\theta_1$ required to align with the angle of each filter tap.
Since all multiplications involve the constant \( \theta_1 \), they can be approximated using only additions and shifts, effectively making the approach \textit{multiplierless}. As shown in Fig.~\ref{root-analysis}c, the penalty stabilizes at 0.5 dB for more than 28 roots. Therefore, we selected 30 roots~(\( \angle\theta_1 = 12^\circ \)).

\begin{figure*}[t!]
\vspace{-6mm}
    \centering
    \includegraphics[width=1\textwidth]{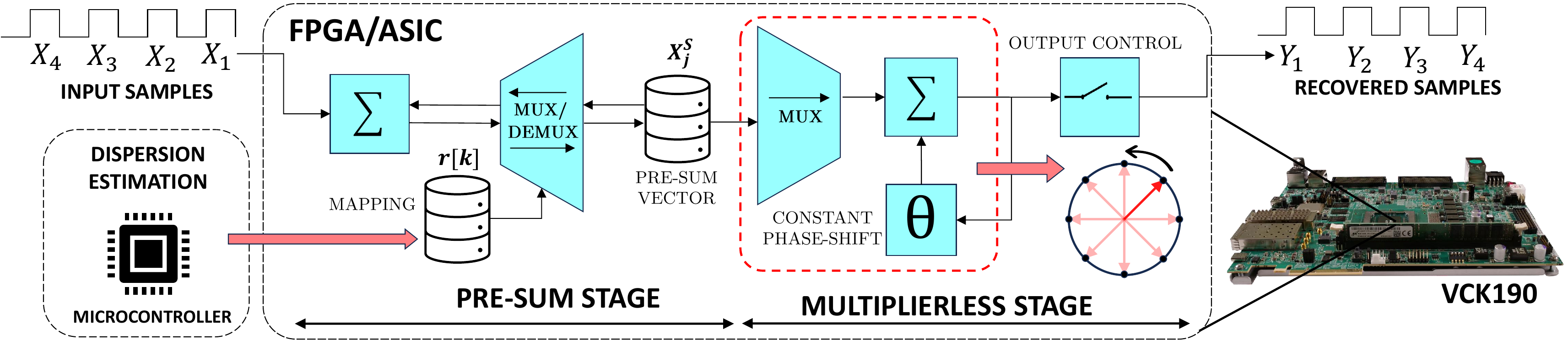}
    \captionsetup{font=footnotesize}
    \caption{Proposed hardware architecture: dispersion estimation is offloaded to an external device to reduce ASIC resource usage. Input samples are accumulated in the pre-sum stage based on the mapping $r[k]$, updated by a microcontroller running a blind or data-aided algorithm. In the multiplierless stage, recursive rotation by $\theta$ in the complex plane is performed using bit-shifts and additions on the pre-summed values $X_j^S$. Both stages are independent of the dispersion scenario, enabling real-world applications without pre-processing.}

    \label{fpga-architecture}
        \vspace{-6mm}
\end{figure*}

\subsection{Flexible Hardware Structure for Different Scenarios and Pre-Processing-Free Features}
To support varying filter sizes, we set a fixed maximum filter size and map unused taps to zero for scenarios with lower dispersion. Fig.~\ref{fpga-architecture} illustrates the proposed hardware architecture. In this design, the pre-summed values, $X_j^S$, are stored in memory and computed by sequentially reading and accumulating the input samples based on the mapping $r[k]$. This approach maintains the pre-sum stage circuitry constant across different transmission conditions.
In the multiplierless stage, the hardware required to perform the multiplication between the pre-summed values  also remains unchanged for different scenarios. Additionally, for the estimation of the mapping $r[k]$, we exploit the slow-changing nature of bulk CDC \cite{savory2008digital}. To further optimize power and area usage in the ASIC, we offload this mapping task to a low-power microcontroller. This enables real-time updates of the mapping $r[k]$, while keeping the ASIC circuitry unchanged and adaptable to various dispersion scenarios.
\begin{figure*}[b!]
    \centering
    \includegraphics[width=1\textwidth]{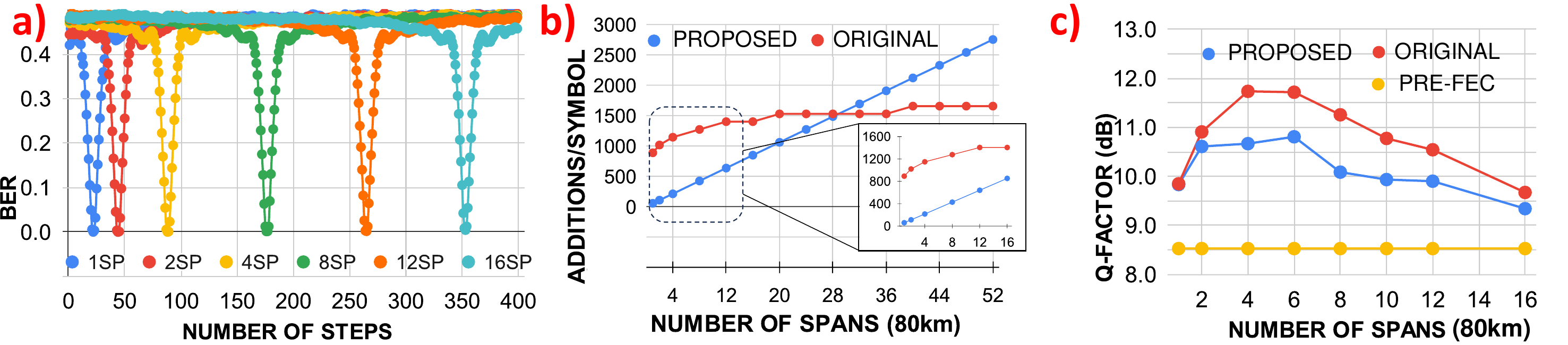}
    \captionsetup{font=footnotesize}
\caption{a) CD-scanning for various fiber lengths, implemented in Python using 5000 symbols for offline performance. b) Computational complexity in generalized additions per symbol, showing gains up to 28 spans, with larger improvements up to 16 spans. c) Performance penalty across fiber lengths, even in worst-case scenarios, staying 1.5 dB above the pre-FEC limit. Simulation parameters: 32 GBaud, single-channel, 16-QAM dual-polarization system at optimal launch power over SSMF with $D$ = 16.8 ps/(nm$\cdot$km), $\gamma$ = 1.2 (W$\cdot$km)$^{-1}$, 0.21 dB/km attenuation, and 4.5 dB EDFA noise figure per 80 km span.}
    \label{plots}
\end{figure*}

We validated this flexible architecture using an adapted CD-scanning method. The angles of the filter coefficients and mappings are obtained by adapting \cite[Eq. 9]{savory2008digital} as:
\(
\phi[k] = \left( \pi/4 + \pi \alpha k^2 \right) \text{mod}(2\pi)
\) and $r[k]~=~(\phi[k]/\angle\theta_1)\text{mod}(N_R)$, where $\alpha = \sqrt{1/(N-1)}$, $N_R$ is the number of roots, $N$ is the required number of taps, and $k$ is the tap index. The value of $N$ is initially set to a small integer number and increased iteratively, with performance being evaluated through the bit error rate (BER) metric after each iteration. The iterations continue until the configuration that yields the lowest BER is found. As shown in Fig.~\ref{plots}a, our adapted CD-scanning method successfully identifies mappings capable of equalizing across 1--16 spans of standard single-mode fiber (SSMF).


\begin{figure*}[t]
\vspace{-10mm}
    \centering
    \hspace{-5mm}
    \includegraphics[width=\textwidth]{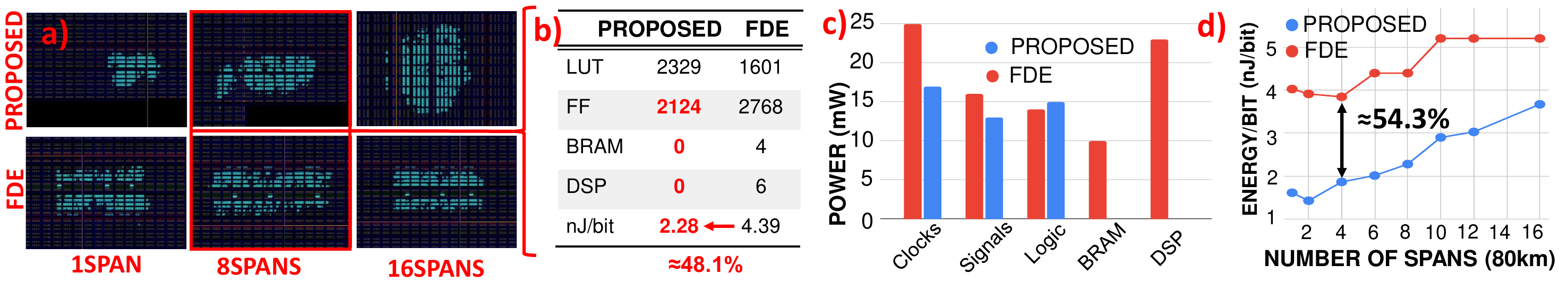}
    \captionsetup{font=footnotesize}
    \caption{a) Chip area usage across fiber lengths, showing minimal area for short to medium distances. b) Hardware resource usage post-RTL synthesis for both equalizers (max reach: 8 spans), showing a 48.1\% energy efficiency gain, with no use of DSP or BRAM and improved FF usage. c) Power distribution breakdown for both equalizers, highlighting energy savings in Clocks, Signals, and particularly BRAM and DSP for 8 spans reach. d) Energy efficiency plot showing that for short fiber lengths, other factors dominate energy consumption, while from 4 spans onward, efficiency is driven by computational complexity, with up to 54.3\% gains.}

    \label{fpga-results}
    \vspace{-6mm}
\end{figure*}

\subsection{Computational Complexity (CC) and Performance Penalty Evaluation}
As a benchmark for comparison, we used the FFT-based Frequency Domain Equalizer (FDE) \cite{xu2010chromatic} . Our proposed method leverages only additions and shifts; therefore, we ignore the costs associated with shifting. To facilitate this comparison, we convert the multiplications in the FDE to additions, under the assumption that each multiplication of two $b$-bit numbers requires $(b-1)$ additions.
The computational complexity of the FDE, using a radix-2 implementation, is given by:
\(
CC_{\text{FDE}}=\frac{4bN_{FFT}(\log_2{N_{FFT}}+1)-2(N_{FFT}+1)}{N_{FFT}-K}
\), where \(N_{FFT}\) is the FFT size and \(K\) represents the overlap. In contrast, our RUE requires only \(2(N-1)\) additions per symbol, with \(N\) being the filter size for RUE. For both equalizers, we set \(b = 16\).
The complexity evaluation is based on the maximum fiber length achievable by both equalizers, employing CD-scanning for RUE and using pre-calculated ideal filter coefficients for FDE. The FDE employs the overlap-save method, processing 2 samples per symbol with a 50\% overlap. We assume the number of taps for both filters corresponds to 60\% of the optimal tap count needed for maximum reach, plus an additional 2 taps, leading to $\log_{2}{N_{FFT}}$ be 6, 7, 8, 9, 9, 10, 10 and 10 for 1, 2, 4, 6, 8, 10, 12 and 16 spans respectively.
As shown in Fig.~\ref{plots}b, our approach exhibits superior computational complexity up to 28 spans (2240 km SSMF), with more pronounced advantages in the 1-16 span (1280 km) range. Furthermore, Fig.~\ref{plots}c demonstrates that even under the worst-case penalty scenario, our solution's performance remains 1.5 dB above the 7\% overhead error-free pre-FEC limit.

\section{FPGA Implementation Results}

We assume that each filter operates as a unitary block, allowing replication for higher baud rates. All analyses were conducted on these blocks implemented on a VCK190 FPGA, simulated in the AMD Vitis IDE at 250 MHz. The FDE utilized AMD's optimized FFT HLS Library for FFT/IFFT operations, along with frequency-domain multiplication and the removal of corrupted samples. Both filters employed 16-bit quantization for taps and signals and were designed for identical throughput of 0.04~$\pm$~6\%~samples/clock across varying fiber lengths. Validation was performed by synthesizing the codes to RTL in Vitis HLS 2023.2 and utilizing the C/RTL Co-simulation tool. The RTL was then implemented in Vivado 2023.2, with energy consumption assessed via Vivado’s Power Estimator, focusing solely on dynamic power. Fig.~\ref{fpga-results}a illustrates that the chip area for the RUE scales better than for FDE for shorter distances. This is primarily due to the Xilinx IP architecture, which leads to a linear increase in latency with FFT size rather than a proportional increase in hardware resources \cite{xilinxfftdatasheet}. In Fig.~\ref{fpga-results}b, the RUE demonstrates significant advantages by avoiding the use of DSPs and BRAMs due to its multiplierless design and efficient memory organization, leading to 48.1\% energy saving compared to the FDE when tested over 8 spans.

Fig.~\ref{fpga-results}c further emphasizes energy savings, showing that the RUE consumes no energy for BRAM and DSP, along with reduced energy use in clock and signal operations. Although there is a slight increase in logic power consumption, this is outweighed by the overall energy efficiency. Additionally, Fig.~\ref{fpga-results}d shows that, at short distances, factors other than algebraic operations drive power consumption; however, from 4 spans onwards, algebraic operations become the dominant factor. The RUE exhibits a linear power consumption, in contrast to the FDE, which shows a stepwise increase due to its escalating computational complexity as the filter reach increases.

\section{Conclusion}
We introduced a novel time-domain equalizer using a symmetric geometric approximation of CDC filter taps to create a multiplierless, pre-processing-free solution that improves energy efficiency and simplifies the ASIC design for coherent transceivers. Our FPGA implementation demonstrated that the RUE achieves energy savings of up to 54.3\% compared to traditional FDE, while maintaining strong performance across various fiber lengths. By eliminating multipliers and pre-processing, the RUE simplifies hardware implementation, enabling real-time adaptability to changing dispersion scenarios, which is essential for modern optical transmission systems.

We acknowledge the support of the EPSRC Progamme Grant TRANSNET.

\bibliographystyle{abbrv}
\bibliography{sample}

\end{document}